\documentclass[MSNbibl,nameyear,dvips]{arxstspdf}
\usepackage{flushend}
\usepackage{stfloats}
\usepackage{graphicx}

\volume{28}
\issue{4}
\pubyear{2013}
\firstpage{466}
\lastpage{486}
\doi{10.1214/13-STS436} 

\makeatletter
\newcommand{\eqref}[1]{(\ref{#1})}
\def\cal{\mathcal}
\renewcommand{\citep}[1]{(\citeauthor{#1}, \citeyear{#1})}
\newcommand{\mbf}[1]{\mathbf{#1}}
\newcommand{\mbv}[1]{\bolds{#1}}

\newcommand{\bfu}{\mathbf{u}}

\newcommand{\reals}{\mathbb{R}}
\makeatother

\begin{document}
\begin{frontmatter}

\title{Modern Statistical Methods in Oceanography: A Hierarchical Perspective}
\runtitle{Hierarchical Statistical Models: Oceanography}

\begin{aug}
\author[A]{\fnms{Christopher K.} \snm{Wikle}\corref{}\ead[label=e1]{wiklec@missouri.edu}},
\author[B]{\fnms{Ralph F.} \snm{Milliff}\ead[label=e2]{Ralph.Milliff@colorado.edu}},
\author[C]{\fnms{Radu} \snm{Herbei}\ead[label=e3]{herbei@stat.osu.edu}}
\and
\author[D]{\fnms{William B.} \snm{Leeds}\ead[label=e4]{leedsw@uchicago.edu}}
\runauthor{Wikle, Milliff, Herbei and Leeds}

\affiliation{University of Missouri, University of Colorado, Ohio State
University
and University of Chicago}

\address[A]{Christopher K. Wikle is Professor, Department of
Statistics, University of Missouri,146 Middlebush Hall, Columbia, Missouri
65203, USA \printead{e1}.}
\address[B]{Ralph F. Milliff is Senior Research Associate, CIRES,
University of Colorado, Boulder, Colorado, USA \printead{e2}.}
\address[C]{Radu Herbei is Assistant Professor, Department of
Statistics, The Ohio State University, Columbus, Ohio, USA \printead{e3}.}
\address[D]{William B. Leeds is Postdoctoral Researcher, Department of
Statistics and Department of Geophysical Sciences, University of
Chicago, Chicago, Illinois, USA \printead{e4}.}

\end{aug}

%
\begin{abstract}
Processes in ocean physics, air--sea interaction and ocean
biogeochemistry span
enormous ranges in spatial and temporal scales, that is, from molecular
to planetary and
from seconds to millennia. Identifying and implementing sustainable
human practices
depend critically on our understandings of key aspects of ocean physics
and ecology
within these scale ranges.
The set of all ocean data is distorted such that three- and
four-dimensional (i.e., time-dependent)
{in situ} data are very sparse, while observations of surface and
upper ocean
properties from space-borne platforms have become abundant in the past
few decades.
Precisions in observations of all types vary as well.
In the face of these challenges, the interface between Statistics and
Oceanography
has proven to be a fruitful area for research and the development of
useful models.
With the recognition of the key importance of identifying, quantifying
and managing uncertainty in data and
models of ocean processes, a hierarchical perspective has become increasingly
productive. As examples, we review a heterogeneous mix of studies from
our own work
demonstrating Bayesian hierarchical model applications in ocean
physics, air--sea interaction,
ocean forecasting and ocean ecosystem models. This review is by no
means exhaustive and
we have endeavored to identify hierarchical modeling work reported by
others across
the broad range of ocean-related topics reported in the statistical literature.
We conclude by noting relevant ocean-statistics problems on the
immediate research horizon,
and some technical challenges they pose, for example, in terms of
nonlinearity, dimensionality
and computing.

\end{abstract}

%

\end{frontmatter}

\section{Introduction}\label{sec1}

The global ocean dominates the iconic image of Earth viewed from space,
leading to the now famous ``blue marble'' descriptor for our planet. The
ocean covers
more than 70\% of the planetary surface and ocean processes are
critical to
life-sustaining (and life-challenging) events and processes occurring
across broad
ranges of temporal and spatial scales.
Understanding issues of ocean resource consumption (e.g., fisheries,
coastal pollution, etc.) lead
to foci on ocean ecosystem dynamics and their coupling to physical processes
(e.g., mixing, transports, upwelling, ice dynamics, etc.).
Understanding the
ocean role in climate dynamics (e.g., sequestration of atmospheric $\mathrm{CO}_2$, absorption
of atmospheric heat, impacts on the hydrologic cycle,
teleconnections,\break
etc.) lead to foci
on massive and complex simulations and forecasts based on equations of
geophysical
fluid dynamics. In all instances, the broad range of scales, the energetic
exchanges across them, and the associated uncertainties drive
innovations that involve
methods of modern statistical modeling.\looseness=1

Oceanography has historically been a ``data poor'' science and the need
to use advanced statistical meth\-odology to perform inference and prediction
has been paramount throughout its development. Although it is the case
that there are too few {in situ}
observations of the ocean to characterize its evolution and its
interaction with marine ecosystems, in an ironic twist,
the discipline also suffers from having an abundance of particular data
types when one factors in the satellite observations that
have become available in the last couple of decades.
The need for statistical collaboration in Oceanography results from
both the situation of not having enough observations in some parts of
the system and having huge amounts of data in other parts of the system.

\subsection{The Physical Ocean}\label{sec1.1}

The physical ocean is governed by basic laws of physics (see, e.g., \cite{vallis2006atmospheric}).
The primitive equations consist of the following: three equations
corresponding to the conservation of momentum
(for the two horizontal and one vertical components of velocity), a
continuity equation representing the conservation of mass,
an equation of state (relating density, pressure, temperature and
salinity), and equations corresponding to the conservation of
temperature and salinity.
There are seven state variables (three velocity components, density,
pressure, temperature and salinity).
This system of equations is nonlinear and exhibits a huge range of
spatial and temporal scales of variability.
Given that many of these scales of variability are not resolved in data
or in deterministic or ``forward'' ocean models,
the equations are typically simplified by scale analysis arguments and
the small scale (turbulent) structures are parameterized. These
parameterizations
involve relationships between the mean of the state variables and their
gradients.
In this way the eddy viscosity and diffusivity terms serve as so-called
``sub-grid scale'' parameterizations, representing the unresolved
processes that are sinks of momentum and heat at the grid scale of a
given model, for example, $O(10)$~km in global ocean models and $O(1)$~km
in regional ocean models.

The ocean system is nonlinearly coupled to the atmosphere across a
broad range of scales. At the largest scales, air--sea fluxes of heat
and fresh water drive mostly vertical or ``thermohaline'' circulations
while the surface shear stress and
wind stress curl drive mostly horizontal or wind-driven gyre
circulations (e.g., \cite{pedlosky1998ocean}). At smaller scales, the
vertical-horizontal separation
breaks down and the ocean response to external forcing and internal
instabilities results in a broadband
spectrum of vigorous eddy circulations (e.g., \cite{mcwilliams2006fundamentals}).
On all scales, the circulation provides the context for ocean processes
affecting other components of the ocean system such as those related to
ocean biology and chemistry.
There are significant nonlinear interactions between the ocean
chemistry, biology and its physical state.

\subsection{Ocean Biogeochemistry}\label{sec1.2}

Ocean biogeochemistry is concerned with the interaction of the biology,
chemistry and geology of the ocean (e.g., \cite{miller2005biological}). This is a very complex system that
contains many interactions across a variety of scales. The system can
be simply illustrated by thinking about the interactions of broad
classes of its components. For example, the presence of nutrients near
the ocean surface, where there is light, allows for the growth of
phytoplankton, which deplete the nutrients as their population expands.
The increased abundance of phytoplankton then provides a food source
for zooplankton, which leads to growth in the zooplankton population.
The consumption of phytoplankton leads to waste products from the
zooplankton that
settles as detritus to the ocean floor. As the zooplankton deplete the
phytoplankton, the zooplankton population decreases due to the lack of
a sufficient food source. Eventually, the detritus at depth is
transferred to the surface through upwelling and mixing, providing the
nutrients that lead to another bloom in phytoplankton, etc.

This simple four component system is a vast oversimplification, as
there are many different species interacting at any one time. More
critically, this lower trophic ecosystem is also coupled to higher
levels of the food
web, for example, with foraging fish predating the zooplankton, which
are in turn predated by higher trophic fish, marine mammals, commercial
fishing, etc.
The chemical component of the cycle is critical in several respects. It
provides a way for carbon to be removed from the atmosphere, as the
phytoplankton remove carbon from the ocean water and are consumed by
the zooplankton. Some of that carbon is contained in the detritus that
sinks to the ocean floor, becoming buried in the sediment and leading
to a (temporary) carbon sink in the global carbon cycle, which is very
important in the context of sequestration of carbon relative to
potential climate change from greenhouse gases. In addition, the
biological cycle in the ocean is closely tied to the distribution of
dissolved oxygen in the water and also influences the distribution of
other chemicals, such as silicon, nitrates and phosphates, for example,
in the shells of diatoms.

\subsection{Uncertainty}\label{sec1.3}

Given the complexities in the ocean system,
it is not surprising that there are numerous sources of uncertainty. First,
although the large-scale equations of motion are in some sense
deterministic, the
scale issues that lead to eddy viscosity/diffusivity parameterizations
are inherently uncertain.
Furthermore, the forms of the linkages between system components (e.g.,
wind stress, heat and moisture fluxes between atmosphere and ocean)
are not known with certainty. The components of traditional
biogeochemical models are even more uncertain,
both in terms of the functional forms and parameters.

The process and parameter uncertainty
is compounded by the inherent data issues in the ocean system.
{In situ} observations of the ocean are quite limited in terms of
spatial and temporal resolution,
and in terms of the variables measured.
For example, it is a painstaking process to measure zooplankton abundance,
often requiring a scientist or technician to literally count critters
through a microscope.
Fortunately, many surface variables can be observed remotely,
particularly through satellite proxies.
In some cases, for example, near surface winds from scatterometers, sea
surface height from altimeters and sea surface temperature (SST) from
radiometers,
the satellite observations are typically quite precise, albeit with
gaps corresponding to orbital geometries, swath widths and fields of view.
In other cases, for example, ocean color as a proxy for phytoplankton,
the observational representation of the process is more uncertain,
at least on fairly short time scales. Thus, a key issue in state
prediction, parameter estimation
and inference is to deal with incomplete observations that vary in
precision and in spatial and temporal support.
This is particularly important when one considers ocean ``data
assimilation,'' that is, the blending of prior information
(e.g., a numerical solution of the deterministic representation of the
ocean state) with observations.

Another traditionally important component of uncertainty in ocean
process modeling corresponds to the selection of reduced-dimensional
representations of the process.
Given the
assumption that much of the larger scale processes in the ocean can be
represented in a lower dimensional manifold, with smaller scales
corresponding to turbulent scales (that certainly may interact with the
larger scale modes, or at least suggest the form of parameterizations
or stochastic noise terms), there has been considerable attention given
to different approaches to obtain the reduced-dimension basis
functions. The choices vary depending on the part of the system being
considered as well as whether one is looking at the system
diagnostically or predictively.

In the context of statistical models used to describe or predict
portions of the ocean system, the nature of the error structures is
important. Given the nonlinearity that is inherent in the system, many
process distributions are not well represented by\break  Gaussian errors. In
addition, in some cases (e.g., biological abundance variables) the
distributions can only have positive support.

\subsection{Statistical Methods}\label{sec1.4}

The ocean and atmospheric sciences have benefitted from a strong
tradition in applying fairly complex statistical methods to deal with
many of the uncertainty issues described above. In particular, general
monographs such as \citet{emery2001data}, \citet{von2002statistical}, and
\citet{wilks2011statistical} provide comprehensive descriptions of
traditional methods used to analyze such data. In addition
to overviews of basic statistical concepts, these books describe
multivariate methods (e.g., principal components---empirical orthogonal
functions,
canonical correlation analysis, discriminant analysis, etc.), spectral
methods (e.g., cross-spectral analysis)
and dynamically-based reduction methods (e.g., principal oscillation patterns)
to facilitate analysis of high-dimensional data that has inherent
dependence in time and space.
This is in addition to more fo\-cused monographs such as
\citeauthor{preisendorfer1988principal}\break
(\citeyear{preisendorfer1988principal}), which gives a comprehensive overview
of\break
eigen-decomposition methods,
and several books and review papers devoted to various aspects of data
assimilation (see Section~\ref{sec3.1}).
Statistical presentations of many of these methods can be found in
\citet
{jolliffe2005principal} and \citet{cressie2011statistics}.

Recognizing the challenges related to uncertainty in the ocean system
and the need to foster more collaborative research between
oceanographers and statisticians,
the U.S. National Research Council\break (NRC) commissioned a panel to write a
report on ``Statistics and \mbox{Physical} Oceanography'' \citep{nrc1994statocean};
see also the accompanying article by \citet{chelton1994physical} and
published comments.
This report contains a very nice review of physical oceanography for
nonoceanographers and outlined the need for research in several key
areas, including the change of support problem and the indirect
nature of satellite observations, non-Gaussian random fields, the
incorporation of Lagrangian and Eulerian data, data assimilation,
inverse modeling,
model/data comparison and feature identification, to name some of the
most prominent.
The report focused on the physical component of the ocean and did not
address biogeochemistry nor many issues of current interest,
such as climate change and reduced-dimensional representations.

\subsection{Paper Outline}\label{sec1.5}

Our goal with this review is to provide an overview of some of the advancements
that have occurred at the interface of Statistics and Oceanography
since the \citet{nrc1994statocean} report.
In particular, we believe strongly that the hierarchical statistical
perspective has played a significant role
in this development and will focus our review from that perspective.
Section~\ref{sec2} presents a brief discussion of hierarchical modeling, both
empirical and Bayesian, with some discussion of the need for
computational tools.
We note that although this paper is in a Statistics journal, we
hope that it will generate interest from both statisticians and oceanographers.
For the same reason that we gave a brief and general overview of
Oceanography above, we will also give a brief and general overview of
hierarchical modeling for those readers with little exposure to these ideas.
In Section~\ref{sec3} we focus in more depth on examples related to data
assimilation and inverse modeling, long lead forecasting
and uncertainty quantification in biogeochemical models.
We will follow this review with a brief discussion of current and
future challenges in Section~\ref{sec4}.

\section{The Hierarchical Modeling Paradigm}\label{sec2}

The idea of hierarchical modeling of scientific processes arose largely
out of \citet{berliner1996hierarchical}, when Mark Berliner was the
director of the Geophysical Statistics Project at the National Center
for Atmospheric Research. The idea, although fundamentally quite
simple, was revolutionary in that it provided a probabilistically
consistent way to
partition uncertainty in systems with complicated data, process and
parameter relationships,
and coincided with the development and popularization of Markov Chain
Monte Carlo (MCMC) methods in Bayesian statistics.
As described below, the key idea is to consider the joint model of
data, process and parameters as three general linked model components,
that is,
the data conditioned on the process and parameters, the process
conditioned on parameters, and the parameters.
These ideas quickly spread into Statistics and subject matter journals
in Climatology, Meteorology, Oceanography and Ecology, to name a few.
This particular perspective on hierarchical modeling is summarized in
several books, including
\citet{clark2007models},
\citet{royle2008hierarchical} and
\citet{cressie2011statistics}.
More traditional Bayesian presentations of hierarchical models can be
found in many books on Bayesian statistics (e.g., \cite{gelman2003bayesian}; \cite{banerjee2003hierarchical}).

Statistical modeling and analysis are about the synthesis of
information. This information may come from expert opinion, physical laws,
previous empirical results or various observations---both direct and
indirect. Consider the case where we have a scientific process of interest,
denoted by $Y$. As an example, say that this process corresponds to the
near-surface north/south and east/west wind components over a portion
of the ocean (i.e., a multivariate spatio-temporal process). We also
have observed data associated with this process, say $Z$, which might
come from a satellite-based scatterometer (i.e., wind component
observations derived from speed and direction that are incomplete in
space and time). We assume that we have parameters associated with the
measurement process, say $\theta_Z$, that might represent differences
in support and representativeness between the satellite observations
and the true wind process at the resolution of interest. In addition,
we assume that there are some parameters, say $\theta_Y$, that describe
the underlying wind process dynamics (i.e., the evolution operator and
innovation covariances that propagate the joint spatial fields of the
wind components through time). Thus, using the total law of
probability, it is natural to write the decomposition of the joint
distribution of the data and process conditioned on the parameters as
%
%
\begin{equation}
[Z, Y | \theta_Z, \theta_Y] = [Z | Y,
\theta_Z] [Y | \theta_Y], \label{eq:ZYhier}
\end{equation}
where $[Z | Y, \theta_Z]$ is the ``data distribution'' (or ``data
model'') and $[Y | \theta_Y]$ is the ``process distribution'' (or
``process model''), and we have assumed conditional independence of the
parameters on the distributions of the right-hand side (RHS) of (\ref
{eq:ZYhier}). Note, we are using brackets ``$[\  ]$'' to refer to a
distribution and the vertical bar ``$  |  $'' to denote ``conditioned
upon.'' Clearly, we could also consider an alternative decomposition in
which $Y$~is conditioned on $Z$ followed by the marginal distribution
of $Z$. However, such a decomposition is less scientific as described below.

In traditional statistics, one might think about the data $Z$ given
some specified distributional form and some associated parameters,
$\theta$ (e.g., corresponding to a spatio-temporal mean and associated
variances and covariances, or their parameterization). In the context
of (\ref{eq:ZYhier}), such a distribution arises from integrating out
the random $Y$ process, yielding the distribution $[Z | \theta\equiv\{
\theta_Z, \theta_Y\}]$.
We are then typically interested in estimating these parameters given
the data. Such estimation (e.g., maximum likelihood estimation) does
not include an explicit representation of a model for the underlying
dynamical wind process, $Y$, but rather includes it implicitly through
the first and second moments (as a consequence of the integration).
In addition, this distribution accounts for the uncertainty that is due
to sampling and measurement.

The question is then why might we be interested in $Y$? First, in many
such applications, one is actually interested in predicting the true,
but unobserved process, $Y$, rather than just accounting for its
(co)variability. Second, given the complexity of most ocean and
atmospheric processes, the multivariate spatiotemporal dependence
structures associated with $Y$ can be very complicated (e.g., nonlinear
in time, nonstationary in space and/or time) and potentially very
high-dimensional. This seriously\break  complicates the likelihood-based
inference, as it puts much of the modeling burden on the realistic
specification of complicated dependence structures. Rather, by focusing
attention on the process $Y$ directly, one can incorporate scientific
insight (e.g., Markovian approximations to mathematical representations
of the process, spatially or time-varying parameters, etc.) and,
critically, disentangle the measurement uncertainty (which can also be
quite complicated) and the process (co)variability and uncertainty.
That is, marginal means and covariances contain potentially complicated
functions of $\theta_Z$ and $\theta_Y$, which can be difficult to
specify without explicitly doing the integration of the random process.
Thus, one is effectively trading the complexity of specifying very
complicated marginal dependence structures with a more scientific
specification of the conditional mean as a random process at the next
level of the hierarchy. This also allows one to focus effort on
modeling the conditional error structure in the data stage, without
having to try to disentangle the measurement uncertainty with the
process (co)variability. Statisticians will recognize this as just a
manifestation of the issue that one faces in traditional mixed-model
analysis where one has a choice of considering a marginal model,
whereby the random effects are integrated out, or a conditional model,
whereby the random effects are predicted and the conditional covariance
structure in the data model is simpler (e.g., \cite{verbeke2009linear}).

Applying Bayes' rule to the hierarchical decomposition in (\ref
{eq:ZYhier}) gives
%
%
\begin{equation}
[Y | Z, \theta_Z, \theta_Y] \propto[Z | Y,
\theta_Z] [Y | \theta_Y], \label{eq:ZYbayes}
\end{equation}
with the normalizing constant obtained by the integral of (\ref
{eq:ZYhier}) with respect to the process $Y$. Note that this implies
that we are updating our knowledge of the process of interest given the
data we have observed. This is the goal of prediction and confirms that
the hierarchical decomposition on the RHS of (\ref{eq:ZYbayes}) is the
more plausible scientific decomposition of the joint distribution of
data and process described above in (\ref{eq:ZYhier}).
Clearly, (\ref{eq:ZYbayes}) assumes that the parameters are known
without uncertainty. This is seldom the case in reality, although one
may have estimates of these parameters from some source and be
comfortable with substituting them into~(\ref{eq:ZYbayes}),
leading to what \citet{cressie2011statistics} refer to as an empirical
hierarchical model (EHM).
Alternatively, and more realistically for most ocean processes, at
least some of the parameters are typically not known,
and we are interested in learning more about them or, at least, would
like to account for their uncertainty. In this case, we consider the
fully-hierarchical or Bayesian hierarchical model (BHM):
%
%
\begin{equation}
[Y, \theta_Y, \theta_Z | Z] \propto[Z | Y,
\theta_Z] [Y | \theta _Y] [\theta_Z,
\theta_Y], \label{eq:ZYbhm}
\end{equation}
where one must specify a prior distribution for the parameters $[\theta
_Z, \theta_Y]$ and we note that the normalizing constant integrates
over the parameters in addition to the process. It is often the case
that we have information to inform these parameter distributions. For
example, in the case of the wind process described above, we have
knowledge about the quality of scatterometer observations of wind
components, and recognize that certain turbulent scaling laws must be
followed, which can be incorporated through restrictions and
informative prior distributions (e.g., \cite{wikle2001spatiotemporal}).

Schematically, it is helpful to think about the RHS of (\ref{eq:ZYbhm})
by using the following representation of Berli\-ner (\citeyear{berliner1996hierarchical}):
%
%
\begin{eqnarray}
\label{eq:BHM} &&[\mbox{data}, \mbox{process}, \mbox{parameters}]
\nonumber
\\
&&\quad = [\mbox{data} | \mbox{process}, \mbox{parameters}]
\\
& &\qquad{} \times [\mbox{process} | \mbox{parameters}] \times[
\mbox{parameters}].
\nonumber
\end{eqnarray}
The strength of this hierarchical representation is that it provides a
probabilistically consistent way to think about modeling complex
systems while quantifying uncertainty. Critically, each of the stages
on the RHS of (\ref{eq:ZYbhm}) or (\ref{eq:BHM}) can be split into
sub-models. For example, multiple data sets with differing supports can
be accommodated with a model such as
%
%
\begin{eqnarray}\label{eq:Z1Z2}
&&\bigl[Z^{(1)}, Z^{(2)} | Y, \theta_{Z^{(1)}},
\theta_{Z^{(2)}} \bigr]
\nonumber
\\[-8pt]
\\[-8pt]
\nonumber
&&\quad = \bigl[Z^{(1)} | Y, \theta_{Z^{(1)}}
\bigr] \bigl[ Z^{(2)} | Y, \theta_{Z^{(2)}} \bigr],
\end{eqnarray}
where $Z^{(1)}$ and $Z^{(2)}$ correspond to data sets (1) and~(2),
respectively, with associated parameters\break $ \theta_{Z^{(1)}}$ and $
\theta_{Z^{(2)}}$. We note that (\ref{eq:Z1Z2}) makes the assumption
that, conditioned on the true process, both data sets are independent.
This is often a reasonable simplifying assumption and greatly
facilitates the combination of \mbox{differing} data sets (although this
assumption should be verified in specific applications). The parameters
and distributional form for the two distributions on the RHS of (\ref
{eq:Z1Z2}) can be quite different, perhaps accommodating differing
types of spatial or temporal support, and/or measurement and sampling
errors. Examples of this in the context of the wind example discussed
previously are data from satellite scatterometers as well as data from
ocean buoys or even weather center analysis winds (e.g., \cite{wikle2001spatiotemporal}).

It is also important to note that the process model component on the
RHS of (\ref{eq:ZYbhm}) or (\ref{eq:BHM}) can also be decomposed into
subcomponents. This could correspond to a hierarchical Markov
decomposition in time, for example, $[Y] = \prod_{t=1}^T [Y_t |
Y_{t-1}][Y_0]$, where $Y = \{Y_0,Y_1,\ldots, Y_T\}$. Or, it could
correspond to a multivariate decomposition, $[Y] = [Y^{(2)} |
Y^{(1)}][Y^{(1)}]$, where $Y = \{ Y^{(1)}, Y^{(2)}\}$. In both cases,
there would also be process model parameters. Clearly, combinations of
these types of distributions and other types of dependencies (e.g.,
spatial, spatiotemporal, etc.) could be considered. In the context of
the wind example, it would be natural for the wind components to be
represented by $Y^{(2)}$ and these could be conditioned on near surface
atmospheric pressure fields, say $Y^{(1)}$, both of which would be
spatiotemporal processes with further conditioning possible.

Last, we recognize that a huge advantage of hierarchical models in
complicated systems is that the parameters are themselves endowed with
potentially quite complex distributions. That is, they exhibit
multivariate dependence between parameters or in terms of space and
time, or may themselves be dependent on exogenous information. Such
complex dependencies in parameters are very difficult to accommodate in
the classical paradigm. As discussed above, in the wind example, it is
critical to specify parameter distributions that adhere to known
empirical and theoretical turbulent scaling properties (e.g., \cite{wikle2001spatiotemporal}).

There is no free lunch! Although the hierarchical modeling paradigm is
extremely powerful, it often comes at a substantial computational cost.
In particular, the normalizing constant in (\ref{eq:ZYbhm}) involves
the integration over all random quantities in the model, which can
correspond to a very high-dimensional integration in many applications.
Given that analytical solutions to these integrals are almost never
available in complex models, one has to resort to numerical methods. In
the fully Bayesian context this is typically some type of Markov Chain
Monte Carlo algorithm (e.g., \cite{robert2004monte}). In the EHM case
represented schematically in (\ref{eq:ZYbayes}), one may be able to
work out alternative computational solutions [e.g.,
expectation-maximization (E--M) or numerical maximum likelihood or
method-of-moments estimation; e.g., see Chapter~7 of \citet
{cressie2011statistics}]. It is critical to note that the complexity of
these calculations often leads to modifications in model structure to
facilitate practical computation.

\section{Hierarchical Modeling and Oceanography}\label{sec3}

There have been many examples of hierarchical modeling in the ocean
sciences since the late 1990s. In this section we briefly review some
of that work at the interface of Statistics and Oceanography. We focus
most attention on two data assimilation examples
(ocean vector winds and ocean tracer state estimation), long lead
forecasting of SST, and uncertainty quantification and assimilation of
biogeochemical models.
For these topics, we present vignettes to illustrate the power of
hierarchical modeling from problems we have worked on.
We also briefly describe some of the work at the interface related to
other important oceanographic problems.

\subsection{Data Assimilation and Inverse Modeling}\label{sec3.1}

\citet{wikle2007bayesian} summarize data assimilation (DA) as an
approach for optimally blending observations with prior information
concerning the system (i.e., mathematical representations of
mechanistic relationships, model output, etc.) to obtain a
distributional characterization of the true state of the system.
Relevant to this overview paper, these concepts originated in the
atmospheric and ocean sciences and there is an extensive literature
describing various methodologies (\cite{daley1991atmospheric};
\cite{ghil1991data};
\cite{bennett2002inverse};
\cite{kalnay2003atmospheric};
\cite{lorenc1986analysis};
\cite{tarantola1987inverse};
\cite{wikle2007bayesian}).
In essence, DA can be thought of as an inverse problem, and one can
derive algorithms from numerous perspectives, including optimal
estimation theory, variational analysis and Bayesian statistics. The
Bayesian approach to DA (e.g., \cite{lorenc1986analysis}; \cite{tarantola1987inverse}) is helpful because
the problem is inherently hierarchical and, thus, it provides a
coherent probabilistic approach that can be used to describe the
various approaches. In addition, if one is willing to consider the BHM
perspective, one can often obtain a more realistic characterization of
uncertainty in various components of the data and mechanistic models
(e.g., \cite{wikle2007bayesian}).\looseness=-1

Most oceanographic processes of interest in the DA context concern
spatial fields that vary with time. A~general dynamical spatiotemporal
model\break  (DSTM) can be written in the BHM paradigm (e.g., \cite{cressie2011statistics}).
The data model is given by
\[
{\mbf Z}_t(\cdot) = {\cal{H}}\bigl({\mbf Y}_t(\cdot),
{\mbv\theta}_d(t), {\mbv\varepsilon}_t\bigr), \quad
t=1,\ldots,T,
\]
where ${\mbf Z}_t(\cdot)$ represents the data at time $t$, ${\mbf
Y}_t(\cdot)$ the associated process,\vadjust{\goodbreak} where the mapping function, ${\cal
{H}}$, may be linear or nonlinear, the error process, ${\mbv\varepsilon
}_t$, may be additive or multiplicative, and the model depends on
parameters given by ${\mbv\theta}_{d}(t)$ that may be spatial or
time-varying. The process model is given by
\[
{\mbf Y}_t(\cdot) = {\cal M}\bigl({\mbf Y}_{t-1}(\cdot), {
\mbv\theta}_p(t), {\mbv\eta}_t\bigr), \quad t=1,2,\ldots,
\]
where the evolution operator ${\cal M}$ may be linear or nonlinear for
the oceanographic process of interest, the error process, ${\mbv\eta
}_t$, may be additive or multiplicative, and the parameters, ${\mbv
\theta}_p(t)$, may be spatial or time-varying. Note that this model is
assumed to be valid beyond the maximum data time-period ($T$), so that
forecasting is appropriate. Finally, the model is completed by the
specification of distributions for the parameters in the previous
stages, $[{\mbv\theta}_d(t)][{\mbv\theta}_p(t)]$,
where we have assumed that the parameters from the two stages are
independent for convenience. Note, this hierarchical framework would
also include a specification of the initial process distribution,
$[{\mbf Y}_0 | {\mbv\theta}_0]$. The sequential modeling perspective
is a natural framework in which to consider the DA problem, as one can
interpret the prior distribution from above as a forecast distribution,
which gets updated given new observations (e.g., see \cite{cressie2011statistics}).

\subsubsection{Ocean surface vector winds}

Surface winds directly transfer momentum to the ocean and surface wind
speed modulates the exchanges of heat and fresh water
to and from the upper ocean as modeled by bulk transfer formulae (e.g., \cite{large06}).
The advent of space-borne
scatterometer instruments in the 1990s provided the first global wind
fields, on daily timescales,
from observations. Prior to the scatterometer era, ocean winds were
inferred from global
weather forecast models and reanalyses (e.g., \cite{hellerman1983normal}) that depended
upon a very sparse global network of {in situ} wind observations from
buoys and ships of opportunity. Reliable resolutions in the
pre-scatterometer wind fields were
limited to ocean basin scale features (e.g., associated with
large-scale high and low pressure systems)
and monthly averages.

The wind fields retrieved from scatterometer observations are not
direct measures
of the wind, but rather the observations are of the roughness imparted
on the ocean by surface capillary waves
in response to
the shear stress vector at the air--sea interface. The amplitudes and
orientations of surface capillary
waves are in equilibrium with the surface shear stress, and these
amplitudes and orientations are
retrievable from measures of how the capillary waves scatter impinging
radar pulses of known frequency and
polarization, that is, so-called microwave backscatter cross-sections;
for details, see \citet{freilich1996}. Radar backscatter cross-sections
are spatially averaged over wind-vector cells and related
to a surface vector wind (SVW) via a geophysical model function. The
SVW retrievals from
scatterometers are accurate to within at least 2~ms$^{-1}$ in speed
and $30^{\circ}$ in direction.
Depending on the sensor system and agency providing the data,
resolutions are on the order
of 12.5--50~km for up to $90\%$ global coverage on daily
timescales. The SVW retrievals occur in swaths
along the polar-orbitting satellite ground track. Swath widths vary by
system from
600--1800~km, that is, between about 20 and 100 wind-vector cells
across a given swath depending on
resolution. Because of the polar orbit (about 14 polar orbits per day) the
swaths overlap at high latitudes and are separated by gaps in coverage
at low latitudes, with the largest
swath gaps occurring at the equator. Again, depending on the system,
the swath gaps at the
equator are on the order of $10$ degrees longitude and can take up to
three days to fill. The SVW from scatterometers
resolve features of the synoptic storms (e.g., fronts, convergences in
rain bands, closed circulations,
etc.) that form, propagate and dissipate every day, over the world ocean.

To fill the gaps in the scatterometer winds, one could make use of the
complete (yet lower resolution) wind fields from the operational
meteorological centers. However, these wind fields have different
properties than the scatterometer observations.
Differences in the true resolutions of weather-center analyses and
scatterometer winds are
efficiently described in terms of surface wind kinetic energy spectra,
that is, kinetic energy as a function
of spatial wavenumber. \citet{wikle1999surface} showed a power-law relation
for surface winds from multiple data sources in the tropical Pacific
and the power law behavior in surface winds
from scatterometry has been documented for the globe \citep
{milliff1999ocean}. \citet{milliff2011ocean} compare kinetic energy
between weather center analyses and SVW retrievals and find that the
kinetic energy drops off unrealistically in weather center analyses at
smaller scales. For example,
at spatial scales on the order of $10^2$~km, the weather center
kinetic energy content is more than an order of magnitude weaker than
the SVW observations.
The goal of a statistical data assimilation is then to blend the
complete, but energy-deficient, weather\vadjust{\goodbreak} center analyses with the
incomplete, yet energy-realistic, SVW in order to provide spatially
complete wind fields at sub-daily intervals while managing the
uncertainties associated with the different data sources and the
blending procedure.

\citet{wikle2001spatiotemporal} implemented a spatiotemporal BHM for
tropical winds in a region of the
equatorial western Pacific. As noted above, the inter-swath gaps in
scatterometer coverage on daily timescales
are largest in the tropics. A BHM formulation to blend weather-center
analyses and scatterometer\break
winds requires space--time properties that account for the greater
intermittency in the scatterometer data.
This is achieved in the process model stage in \citet
{wikle2001spatiotemporal}, which is posed in terms of three
scientifically-motivated components as
%
%
\begin{equation}
{\mbf Y}_t = {\mbv\mu}^u +{\mbv\Phi}^{(1)}{\mbv
\alpha}_t^u + {\mbv \Phi}^{(2)}{\mbv
\beta}_t^u, \label{tropprocess}
\end{equation}
where ${\mbf Y}_t$ is a vector of spatially-indexed zonal (east--west)
wind velocity components (typically denoted by~$u$). Analogous terms
apply for
the meridional velocity (${v}$). Here, ${\mbv\mu}^u$ is the mean zonal
wind process that
accounts for the prevailing wind and its variance in the tropical domain,
${\mbv\Phi}^{(1)}$ is a basis function matrix corresponding to the
large-scale modes of the equatorial beta-plane approximation (i.e., a
linear, thin fluid approximation)
to the momentum equations, and ${\mbv\Phi}^{(2)}$ are nested wavelet
basis functions used to model fine-scale winds
according to observed kinetic energy-wavenumber properties for the
region. The use of scientifically-motivated basis functions is crucial
here, as we have information from the thin-fluid analytical
approximation to the wind field that is appropriate over the tropics.
In particular, the leading equatorial normal modes [${\mbv\Phi
}^{(1)}$] are large-scale wave signals in the tropics,
that is, westward propagating Rossby waves, eastward Kelvin waves and
mixed Rossby--Gravity waves (e.g., \cite{matsuno1966quasi}). Based on
data analyses (e.g.,\break  \cite{wheeler1999convectively}), the equatorial
normal mode basis functions were limited to the leading modes defined
by two zonal wavenumbers and four wavenumbers in the meridional
direction \citep{wikle2001spatiotemporal}. These are sufficient to
support the important basin-scale Kelvin, Rossby and mixed
Rossby--Gravity waves that can be used to describe much of the tropical
dynamics within the Pacific basin.
The amplitude coefficients, ${\mbv\alpha}^u_t$ and ${\mbv\beta}^u_t$,
are treated as random variables in the BHM, with first-order Markov
models. The priors on the parameters associated with these models
correspond to time series of the theoretical amplitudes for each mode
for the ${\mbv\alpha}^u_t$ coefficients and the slopes of the kinetic
energy spectrum for the ${\mbv\beta}^u_t$ coefficients as obtained
from \citet{wikle1999surface}. Data stage inputs were obtained from
NSCAT SVW retrievals and from sea level pressure and surface winds
taken from reanalyses of the National Centers for Environmental
Protection (NCEP).

\begin{figure}

\includegraphics{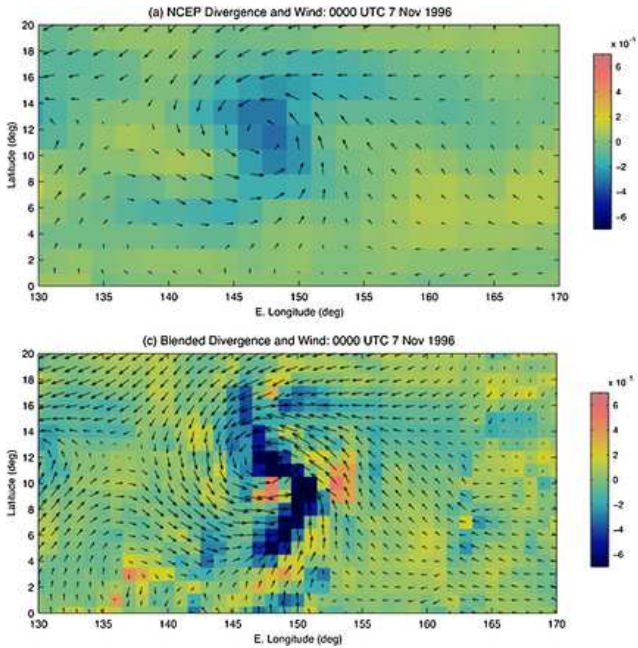}

\caption{Tropical surface wind BHM after Wikle et~al. (\citeyear{wikle2001spatiotemporal}). Top panel is
the NCEP reanalysis at the time tropical cyclone Dale occupied the BHM
domain. Color contours denote convergence (blues) and divergence (reds) in the surface wind field.
Bottom panel is the posterior mean wind field (vectors) and convergence/divergence
(colors) map for tropical cyclone Dale from the
tropical wind BHM. Strong convergences in rainband structures spiraling
into the tropical cyclone center coincide with
coldest cloud-top temperatures from independent satellite observations.}\label{2paneltrop}
\end{figure}

The posterior mean surface wind and surface convergence/divergence
exhibited variability on scales
not achievable in the NCEP reanalyses. This is particularly evident in
a comparison of the NCEP
reanalysis wind field with the posterior mean winds from the tropical
wind BHM for the period when
tropical cyclone Dale crossed the study area (Figure~\ref{2paneltrop}).
The posterior mean
winds are organized into strong convergence regions that spiral into
the center of the tropical cyclone.
These are consistent with rain band features of tropical cyclones and
this was confirmed by comparing the posterior
mean winds with cloud-top temperatures from an independent satellite
observation nearly
coincident with the snapshot from the posterior distribution of the BHM
(see \cite{wikle2001spatiotemporal}).

Uncertainty management properties of SVW BHM based on mechanisitic
models (i.e., leading
order terms and/or approximations of the primitive equations) are
particularly relevant in
ocean forecast settings.
The Mediterranean Forecast System (MFS) is an operational system
producing 10-day forecasts for
upper ocean fields every day. The MFS ocean forecast models resolve
synoptic variability in the upper ocean.
The uncertain parts of the forecast fields are at ocean mesoscales,
that is, hourly and 10--50~km scales.
These are the scales of upper ocean hydrodynamic instabilities driven
by the surface wind.
So, modeling uncertainty in the surface wind field is a useful means of
quantifying uncertainty in the MFS ocean forecasts on the scales that
are most
important to daily users.

\citet{milliff2011ocean} describe the details of the SVW BHM for MFS,
and \citet{pinardi2011ocean}
review the impacts of BHM SVW fields in an ensemble forecast
methodology built around realizations
from the posterior distribution for SVW from the BHM. The mechanistic
process model in \citet{milliff2011ocean}
involves the leading-order terms in a Rayleigh Friction equation
approximation at synoptic scales
and, again, a~nested wavelet basis model to represent turbulent closure
at the finest spatial scales. Thus, again it is critical to incorporate
scientific information into the specification of the process rather
than try to model such behavior through the marginal covariance
structure. This then also allows specification of the measurement
uncertainties associated with the data conditioned upon the true
process. Specifically, data stage inputs are obtained
from the QuikSCAT scatterometer SVW retrievals and from weather center
sea level pressure and surface wind fields.

\begin{figure}

\includegraphics{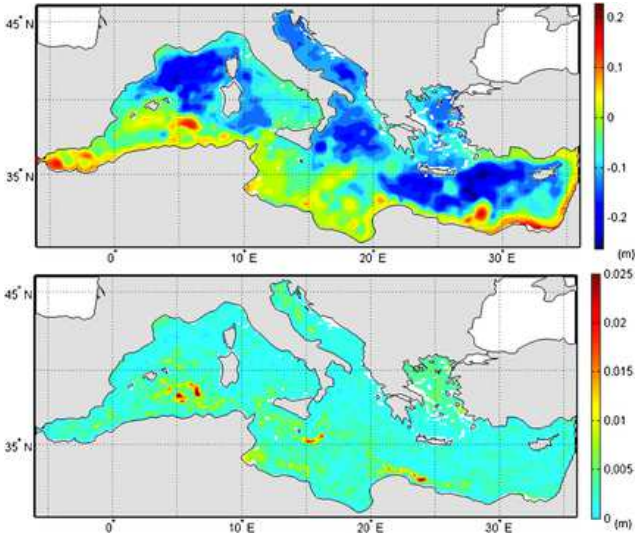}

\caption{Forecast initial condition mean (top) and spread (bottom) in
the sea-surface height field computed from
10 ensemble members, each driven
by a realization from the posterior distribution for the SVW BHM
(Milliff et~al., \citeyear{milliff2011ocean}).
Ensemble spread is usefully localized in the most uncertain regions of
the domain where wind forcing is driving
hydrodynamic instabilities and a local pulse in the mesoscale eddy
field (Pinardi et~al., \citeyear{pinardi2011ocean}).}
\label{2panelmed}
\end{figure}

\begin{figure*}[b]

\includegraphics{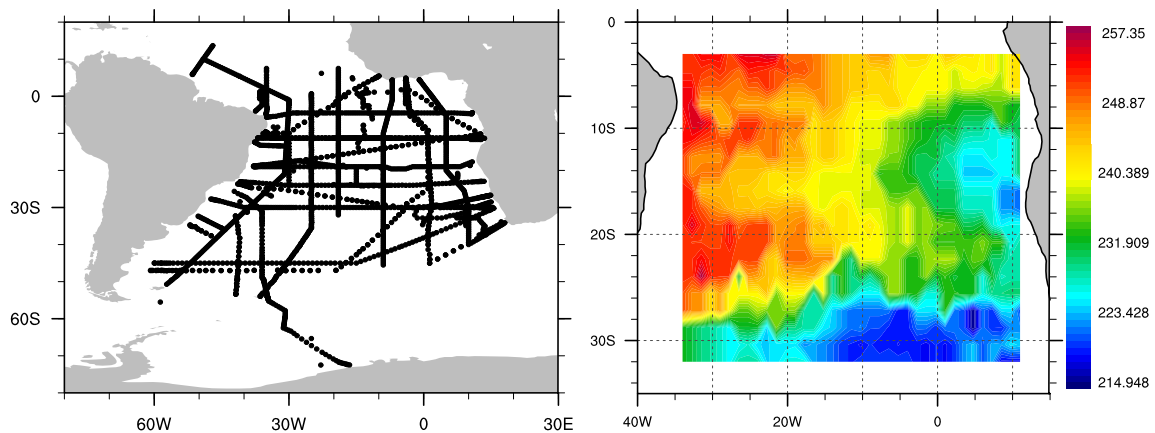}

\caption{Left: World Ocean Circulation Experiment era cruise tracks in
the South Atlantic Ocean. Right: Interpolated oxygen measurements in a
deep neutral layer of the South Atlantic Ocean.}
\label{fig1}
\end{figure*}

Ten realizations of the posterior distribution for SVW were used to drive
ensemble data assimilation and ensemble forecasts in the MFS. Figure~\ref{2panelmed}
depicts the data assimilation system response in sea-surface
height (SSH) at the forecast initial condition time. The mean SSH
initial condition (Figure~\ref{2panelmed}, top) reflects the
accurate spatial scales for MFS forecasts where synoptic variability
overlies the general circulation
patterns for the Mediterranean Sea. Sub-basin scale cyclonic gyres are
represented by blue
(depressed) SSH and anticyclones by reds. The gradients between blue
and red signals are
proportional to surface current speeds. The spread in the SSH initial
condition (Figure~\ref{2panelmed}, bottom)
demonstrates the value of managing uncertainty in the SVW. The largest
amplitude signals in the spread are
localized in a few places only---where surface wind is driving
hydrodynamic instabilities and
pulsing the energy in ocean mesoscale response.

\subsubsection{Inversion of oceanographic tracer data}

Inferring the structure of the circulation in the world's oceans is a
significant part of our quest for understanding the climate. Direct
measurements of the circulation are difficult to obtain; however, in
the past decades, scientists have collected large amounts of
hydrographic data (hydrostatic pressure, temperature, salinity, oxygen
concentration, etc.). Such data have been used to produce
climatological maps that exhibit the large scale structure (Lozier, Owens and Curry \citeyear{lozi1995}).
Based on these maps and data, one can infer some
constituents of the ocean circulation, such as boundary currents,
wind-driven gyres and abyssal interior flow. Yet, hydrographic data are
typically available at very sparse locations in space and time. For
example, the World Ocean Circulation Experiment lasted from 1990 until
1998. In Figure~\ref{fig1} we illustrate these data that were available
for the South Atlantic Ocean. Consequently, statistical smoothing
methods are required to provide estimates at space--time locations of
interest. To that end, one would typically combine data sets collected
at different times. This seems reasonable only under the assumption
that the distribution of dynamically-passive tracers is representative
of a mean circulation and that such tracers are sufficiently stable on
a fixed neutral density layer.

A classical approach to inverting hydrographic data into circulation
structure is the \textit{box inverse method} \citep{wuns1996}. Starting
with the thermal wind equations, the problem reduces to estimation of a
reference velocity field. This is achieved via a large system of linear
equations expressing conservation of mass within a collection of
connected boxes. Due to the size of the problem, a high-dimensional
regression approach is subsequently employed. Traditionally, ridge
regression estimators have proved useful, however, modern
statistical/machine learning methods can provide more effective
alternatives. Other approaches are based on more complex inverse
modeling and nonlinear optimization (\cite{pail1997};
\cite{zika2010};
\cite{zhan1992};
\cite{wuns1994}). The inverse problem described
here is ill-posed and some type of regularization is required \citep
{kirs1996}. We describe the Bayesian inversion approach of \citet{Metal05}
(see also \cite{herb2008}), which connects the data (tracer concentrations) to
parameters (ocean circulation) using a~system of partial differential
equations. In this case, regularization is imposed through a prior
distribution over the parameters. The \textit{solution} is the posterior
distribution. It can be used to select ``representative'' values and
the associated uncertainty for any aspect of the circulation.

Formally, let $C=C(x,y)$ denote the concentration of a tracer of
interest at a location $(x,y)$ in a rectangular domain $\Omega\subset
\reals^2$ representing the layer of the ocean being studied.
The problem is to estimate the (horizontal) water velocities and
diffusion coefficients based on noisy measurements of $C$ at a sparse
set of locations in $\Omega$. In the right panel of Figure~\ref{fig1}
we display an interpolated map of oxygen concentration for a $2000$~m
deep layer in the South Atlantic Ocean. The connection between the
tracer concentration $C$ and the velocities and diffusion coefficients
is modeled by the steady-state advection-diffusion equation
%
%
\begin{equation}
\bfu\cdot\nabla C = \nabla\cdot(K\nabla C) + Q_C,\quad (x,y)\in
\Omega, \label{ad}
\end{equation}
where $\bfu=(u,v)$ is the horizontal water velocity, the diagonal
diffusivity matrix $K = \operatorname{diag}(\kappa^{(x)}, \kappa^{(y)})$ does not vary
with location, and the sink term $Q_C = -\lambda C$ is present only
when $C$ represents oxygen concentration. Equation \eqref{ad} is
augmented with Dirichlet boundary conditions $C = C_{\partial\Omega}$
when $(x,y)\in\partial\Omega$, where $\partial\Omega$ denotes the
boundary of $\Omega$.

The statistical model assumes additive observational error
\[
C_j^{\mathrm{obs}}(x_i,y_i) =
C_j(x_i,y_i) + \varepsilon_j(x_i,y_i).
\]
Here, $j = 1,\ldots, n_C$ indexes a particular tracer and $i=1,\ldots,
N_j$ indexes a spatial location where data for tracer $j$ are
available. The underlying tracer concentration $C(x,y)$ is obtained as
a solution of the advection diffusion equation \eqref{ad}. As this
solution is not available in closed form, one uses a numerical
(grid-based) approximation. We collect all quantities of interest
(velocities, diffusion coefficients, boundary values) in a
high-dimensional vector ${\mbv\gamma}$. Under the \mbox{assumption} that the
measurement errors $\varepsilon(\cdot)$ are unbiased Gaussian variables
with constant (yet tracer-dependent) variance, the posterior
distribution is written as
%
%
\begin{eqnarray}\label{post0}
&&\pi\bigl({\mbv\gamma} | C^{\mathrm{obs}}\bigr)
\nonumber
\\
&&\quad \propto\prod
_{j=1}^{n_C}\prod_{i=1}^{N_j}
\exp\biggl\{ -\frac{(C({\mbv\gamma}; x_i, y_i)-C^{\mathrm{obs}}(x_i, y_i))^2}{2\sigma^2_j }\biggr\}\\
&&\hspace*{52pt}{}\cdot\pi({\mbv\gamma}),\nonumber
\end{eqnarray}
where $\pi({\mbv\gamma})$ represents the selected prior distribution.
The posterior probability model \eqref{post0} is explored via Monte
Carlo methods (e.g., \cite{robert2004monte}).

The ill-posedness mentioned above is resolved by specifying a proper
prior distribution $\pi({\mbv\gamma})$. This implies that the
posterior distribution is proper. However, an efficient MCMC approach
is still required. In addition, one can design specific sampling
strategies as described in \citet{Metal05} and \citet{herb2008}.
It is important to understand that each component described above
(physical model, prior distribution, likelihood function, data) plays a
crucial role in determining the solution. Under the Bayesian approach,
a sensitivity analysis, although possible, is hampered by the immense
computational cost associated with the MCMC sampler. In addition, the
under-determina\-tion problem is present in this case. There are\break
(roughly) 300 data points, while there are thousands of parameters to
estimate (velocities, diffusions,\break boundary values). In the results
given in \citet{Metal05} and \citet{herb2008}, not all estimated velocities are
significantly different from zero, which is the prior mean. However,
the data are informative about the large-scale features (zonal jets,
gyres). The posterior mean velocities are compared with velocities
determined from float data \citep{hogg1999}, showing a comforting
consistency (\cite{Metal05} and \cite{herb2008}).

\subsection{Long Lead Forecasting: SST}\label{sec3.2}

Tropical Pacific SST exhibits some of the most important variability on
inter-annual time scales for the ocean/atmosphere system (e.g., see the overview in \cite{philander1990nino}). This variation arises from
complicated interactions, across a large range of spatiotemporal
scales, between the ocean and the atmosphere. The most prominent signal
on these time scales is the El Ni\~no-Southern Oscillation (ENSO)
phenomenon. This is the well-known quasi-periodic (3--5 year period)
warming (El Ni\~no) and cooling (La Ni\~na) in the tropical Pacific.
These warming and cooling events lead to dramatic effects in weather
across the globe due to teleconnections with the global atmospheric circulation.
Because of these significant weather related impacts (e.g., droughts,\break
floods, etc.), it is critical to be able to forecast several months in
advance the possible development and transition of these events.
Increasingly, such\break  ``long lead'' forecasts have shown useful skill and
correspond to one of the few situations in ocean science where a purely
statistical forecast methodology is competitive with, and in many cases
better than, equivalent deterministic model forecasts
(\cite{barnston1999predictive};
\cite{jan2005did}).

At typical spatial resolutions, there can easily be several thousand
gridded spatial locations corre-\break sponding to the tropical Pacific region
of forecasting interest. Complicated spatiotemporal statistical models
are difficult or impossible to implement at these dimensions. For this
reason, statistically-based models for tropical SST have been
``spectral'' in the sense that they are based on coefficients
associated with a projection of the SST fields on spatial basis
functions. The associated projection coefficients that are evolved are
typically a reduced set, usually corresponding to larger modes. That
is, let ${\mbf Y}_t$ correspond to an $n$-dimensional vector of the
true SST process at $n$ spatial locations and time $t$ and consider the
decomposition of this process vector
%
%
\begin{equation}
{\mbf Y}_t = {\mbv\Phi}^{(1)} {\mbv\alpha}_t +
{\mbv\Phi}^{(2)} {\mbv\beta}_t, \label{eq:Yphiphi}
\end{equation}
where ${\mbv\Phi}^{(i)}, i=1,2$, is an $n \times p_i$ matrix of
spatial basis functions, and ${\mbv\alpha}_t$ and ${\mbv\beta}_t$ are
the associated expansion coefficients. A first-order linear Markov
assumption on the evolution of the coefficients ${\mbv\alpha}_t$, for
example, ${\mbv\alpha}_{t+\tau} = {\mbf M}{\mbv\alpha}_{t} + {\mbv
\eta}_{t+\tau}$, for appropriate time increment, $\tau$, and with
Gaussian errors, ${\mbv\eta}_t \sim \operatorname{Gau}({\mbf0}, {\mbf Q})$, was
shown in the early 1990s to be a model with reasonable skill at
long-lead forecasting (e.g., \cite{penland1993prediction}). As with
the tropical wind case presented previously, the specification of the
process in terms of two basis sets is critical from a scientific
perspective. In particular, it is thought that the active dynamical
process is driven by a lower dimensional manifold (corresponding to the
first set of basis functions) and, yet, the residual
spatially-dependent structures captured by the second set of basis
functions remain an important source of variability.

Although dimension-reduced linear Markov models showed reasonable
skill, the ENSO phenomenon is better characterized as a nonlinear
process (e.g., \cite{hoerling1997nino}; \cite{burgers1999normal}). Nonlinear
statistical models typically are better at capturing the magnitude of
the predicted El Ni\~no and La Ni\~na events
(\cite{tangang1998forecasting};
\cite{berliner2000long};
\cite{tang2000skill};
\cite{timmermann2001empirical};
\cite{kondrashov2005hierarchy}).
Most nonlinear stochastic methods have not successfully characterized
the uncertainty in the forecasts. An exception to this is the model of
\citet{berliner2000long}, who use a dimension-reduced threshold Markov
model with certain components governed by the onset of intra-seasonal
oscillations (so-called ``westerly wind bursts'').

There are several key components of the model in \citet
{berliner2000long} that illustrate the power of hierarchical modeling.
First, the data model is constructed
%
%
\begin{equation}
{\mbf Z}_t = {\mbv\Phi}^{(1)} {\mbv\alpha}_t +
{\mbv\varepsilon}_t,\quad  {\mbv\varepsilon}_t \sim \operatorname{Gau}({
\mbf0},{\mbf R}),
\end{equation}
where the data vector, ${\mbf Z}_t$, is very high-dimensional (say, $n
\times1$), and ${\mbv\Phi}^{(1)}$ is a reduced-dimension set of
spatial basis functions as described above (an $n \times p_1$ matrix).
It is important to note that the remaining basis functions (e.g.,
${\mbv\Phi}^{(2)}$ from above) associated with the basis expansion are
used to parameterize the observational spatial covariance matrix,
${\mbf R}$, which now accounts for both observation error and the
truncation. The assumption is that the active dynamics exist on the
lower dimensional manifold represented by ${\mbv\Phi}^{(1)}$ and, like
in typical turbulence parameterizations, the small scale structures
accounted for by the ${\mbv\Phi}^{(2)}$ portion of the expansion are
associated with nonpredictive covariability.

Critically, the evolution of the dynamical process is nonlinear, but
conditionally linear. Specifically,
%
%
\begin{equation}
\qquad{\mbv\alpha}_{t+\tau} = {\mbv\mu}_t + {\mbf M}_t
{\mbv\alpha}_t + {\mbv\eta}_{t+\tau}, \quad {\mbv
\eta}_t \sim \operatorname{Gau}({\mbf0},{\mbf Q}),
\end{equation}
where prior distributions are also specified for $\{{\mbv\alpha}_t\dvtx\break t=1,\ldots, \tau\}$. The important modeling assumption is that ${\mbf
M}_t$ and ${\mbv\mu}_t$ depend on both the current (time $t$) and
future (time $t+\tau$) climate ``regimes,'' that is, ${\mbf M}_t =
{\mbf M}(I_t, J_t)$ and ${\mbv\mu}_t = {\mbv\mu}(I_t,J_t)$, where
$I_t$ classifies the current SST regime as ``cool,'' ``normal'' or
``warm,'' and $J_t$ anticipates one of these three regimes at time
$t+\tau$. Specifically, $I_t$ is based on the Southern Oscillation
Index and $J_t$ is based on a latent variable that is modeled,
hierarchically, in terms of the east-west component of the near surface
wind in the Western Pacific ocean (for details,
see \cite{berliner2000long}). The point is that this leads to nine
potential mean states and nine potential dynamical operators,
accommodating the nonlinear transition between ENSO phases.

Because this model was developed in a Bayesian hierarchical framework,
it was able to capture the uncertainty characterized by the data,
process and parameters. However, despite the fact that this model used
physical notions (i.e., westerly wind bursts) in the lower stages of
the model hierarchy, it was unable to directly account for quadratic
interactions across spatial scales. As mentioned previously,
nonlinearity is important in most atmosphere and ocean processes.
\citet
{kondrashov2005hierarchy} demonstrate the effectiveness of a quadratic
nonlinear model for long-lead prediction of ENSO from a classical
regression perspective. \citet{wikle2010general} demonstrate the
implementation of a quadratic nonlinear model in the context of a
Bayesian hierarchical framework in which the random nonlinear process
is ``hidden'' and parameters are random as well. They illustrate that
simple arguments from turbulence scaling support the form of this model
and further suggest a dimension reduction strategy.

As defined in \citet{wikle2010general}, general quadratic nonlinearity
(GQN) with respect to the ${\mbv\alpha}_t$ process can be written
%
%
\begin{eqnarray}\label{eq:GQN}
&&\alpha_{t+\tau}(i)\nonumber\\
&&\quad= \sum_{j=1}^p
m^L_{ij} \alpha_{t}(j)
\\
&&\qquad{} + \sum
_{k=1}^p \sum_{l=1}^p
m^Q_{i,kl} \alpha_{t}(k) g\bigl(
\alpha_{t}(l);{\mbv\theta}_g\bigr) + \eta_{t+\tau}(i)\nonumber
\end{eqnarray}
for $i =1,\ldots,p$, where $g(\cdot)$ is some transformation of ${\mbv
\alpha}_{t}$ that depends on parameters ${\mbv\theta}_g$ and gives
the process more generality than the simple dyadic interactions alone.
This framework is exceptionally flexible in accommodating many
real-world mechanistic processes, but it comes at the cost of a curse
of dimensionality in the parameter space; that is, there are $O(p^3)$
parameters corresponding to the nonlinear coefficients ($m^Q_{i,kl}$),
in addition to the linear coefficients ($m^L_{ij}$) and~${\mbv\theta}_g$.

One can use scale analysis to help with the dimensionality concerns,
and also to motivate components of the hierarchical model.
Specifically, as before, assume we can decompose the spectral
coefficients into large-scale components (${\mbv\alpha}_t$) and
small-scale coefficients (${\mbv\beta}_t$) corresponding to the
spatial basis functions ${\mbv\Phi}^{(1)}$ and ${\mbv\Phi}^{(2)}$
discussed above. Consider all of the possible dyadic interactions of
the elements of this vector---that is, there are
small-scale/small-scale, small-scale/large-scale and
large-scale/large-scale interactions. Loosely motivated by ``Reynolds
averaging'' in turbulence theory (e.g., \cite{holton2004dynamic}),
\citet{wikle2010general} suggest considering the large-scale/large-scale
pairwise interactions explicitly, with the small-scale/small-scale
interactions contributing to a correlated \mbox{dependence} structure in the
additive error, and the small-scale/large-scale interactions
corresponding to the linear term in the large-scale coefficients with
random parameters (i.e., the small-scale coefficients play the role of
random coefficients in the \mbox{interaction}). Such an argument leads to a
quadratic nonlinear model on the large-scale coefficients, which can be
written in matrix form as
%
%
\begin{eqnarray}\label{eq:GQNvec}
{\mbv\alpha}_{t+\tau} = {\mbf M}_L{\mbv\alpha}_{t}
+ \bigl({\mbf I}_{p_1} \otimes{\mbv\alpha}'_t
\bigr) {\mbf M}_Q {\mbv\alpha}_t + {\mbv\eta
}_{t+\tau},
\nonumber
\\[-8pt]
\\[-8pt]
\eqntext{{\mbv\eta}_t \sim \operatorname{Gau}({\mbf0},{\mbf Q}).}
\end{eqnarray}
Prior distributions are then given to the parameters in ${\mbf M}_L$
and ${\mbf M}_Q$ as well as the covariance matrix ${\mbf Q}$ (e.g., \cite{wikle2010general}). This approach was shown to quite
reasonably account for the uncertainties so that the prediction error
bounds covered the extreme ENSO events, even when the forecasts did not
adequately capture the true magnitude.

Critically, these quadratic nonlinear implementations still suffer from
a fairly high curse of dimensionality in the parameters. In this sense,
various model reduction approaches have to be implemented that
necessarily fail to account for as much model uncertainty as is likely
present. It is difficult to know {a priori} which quadratic
interactions are important. To alleviate this curse of dimensionality
and to provide a natural framework in which to ``average'' across the
various model specifications, \citet{wikle2011polynomial} employed a
stochastic search variable selection methodology (e.g.,
\citeauthor{george1993variable}, \citeyear{george1993variable,george1997approaches}).

In general, although the GQN statistical models are very flexible in
representing oceanographic processes, these models can easily
experience finite-time blow up (i.e., explosive growth) when fit to
data \citep{majda2010fundamental}. This is seldom an issue when one is
using these models for data assimilation, given the presence of
observations to act as a control (e.g., \cite{leeds2012modeling}),
nor is it typically a problem when one is only forecasting out one time
step (e.g., as in the SST example). It can be a problem when multiple
time steps are forecast that require some form of constraint, either
statistical or physical \citep{majda2012physics}.

\subsection{Biogeochemical Models}\label{sec3.3}

Analysis of marine ecosystem dynamics involves various sources of
uncertainty in the observations, the underlying scientific process and
the parameters that describe the process dynamics. Critically, it is
often the case that the observation errors are non-Gaussian and that
the process being modeled is nonlinear, that is, there is an explicit
system of coupled nonlinear differential equations that describe the
complex ecosystem dynamics. As a result of these uncertainties and
complexities, the BHM framework is natural, but can be difficult to implement.

Soon after the introduction of MCMC methods into Bayesian computation,
the BHM approach was used in data assimilation for marine ecosystem
models. Anticipating the forthcoming increase in re-\break motely-sensed ocean
color observations, \citet{harmon1997markov} implemented an MCMC-based
sampling protocol that explored the ability to recover various model
parameters in a seven-compartment marine ecosystem model with and
without added model noise. They noted the computational difficulties
required to adequately account for correlation in these parameters.
While identifying correlation in only ten parameters may no longer be a
computational issue, there are still issues related to the sampling
methodology that adequately generates reasonable block proposals for
the entire parameter vector. Adaptive Metropolis--Hastings algorithms,
which update the covariance of the proposal distribution, may be useful
when trying to generate adequate proposals for a nonlinear marine
ecosystem model (see, e.g., \cite{haario2001adaptive}). Current
research has taken into account advances in sampling methodology,
including the following: sequential importance resampling, particle
filters, ensemble Kalman filters and state-augmentation approaches
(\citeauthor{evensen1994sequential},
\citeyear{evensen1994sequential,evensen2009data};\break
\citeauthor{dowd2006sequential}, \citeyear{dowd2006sequential,dowd2007bayesian,dowd2011estimating};
\cite{parslow2013bayesian};
\cite{stroud2010ensemble}).

An innovative practice in statistical modeling of ocean ecosystems is
the inclusion of information from deterministic, mechanistic models
into the statistical framework, typically in the process stage of a
BHM. However, the necessary estimation procedures often require
iteratively running the mechanistic model, which poses a problem when
the model is computationally expensive and can only be run a very
limited number of times. In certain situations where the computer model
is too computationally expensive to run a sufficient number of times
for the desired analysis, statistical surrogates (i.e., emulators) are
used. In its simplest form, an emulator is simply the resulting
estimated statistical model when computer model output is used as data.
Then, this model is used to predict the output of the computer model
under untried input settings (e.g., initial conditions, parameter
values, forcings).

Traditionally, so-called ``computer model'' emulation has been done
using Gaussian Process (GP) models (e.g., \cite{sacks1989design}; \cite{currin1991bayesian}). In our case, these
``computer models'' are deterministic ocean forward models. GP
emulators are related to spatial GPs, which use a correlation function
such that the model output is more highly correlated for inputs that
are ``nearer'' to one another in a given sense than those that are
``farther apart.'' However, because ocean ecosystem models are
nonlinear, a GP emulator of the joint distribution of the output may be
inappropriate (as a nonlinear process cannot be specified by only two
moments). In this case, one could consider the use of a dynamic GP
emulator, which considers the value of the process at the current time
step (i.e., the initial conditions) as an input to the ocean forward
model (and GP emulator). This offers several benefits over the
traditional GP emulation approach (e.g., \cite{conti2009gaussian};
\cite{liu2009dynamic}).

Rather than modeling the dynamics through a covariance function in a
GP, it may be more appropriate to model the output using first-order
characteristics (as the computer models themselves are written). The
use of so-called ``first-order emulators'' has appeared in several
applications, for example,
\citet{van2007fast} and \citet{frolov2009fast} use neural networks and
\citet{hooten2011assessing} use random forests, a~machine learning
algorithm. There is also the potential to use other nonlinear
statistical models, polynomial chaos expansions
(Mattern, Fennel and Dowd, \citeyear{mattern2012estimating}) or GP models in a dynamical context
\citep{margvelashvili2012sequential}. These approaches use
dimension-reduction techniques to overcome the curse of dimensionality,
with the emulator describing the dynamics of the reduced-dimensional
process. Most of these implementations use nonparametric approaches.
Alternatively, \citet{leeds2012emulator} use an emulator based on a
reduced-rank multivariate quadratic nonlinear statistical model as
described above.

Related to the use of emulators for dynamic models,
there is also a growing body of work considering the use of emulators
for spatial or spatiotemporal
forward model output. Marine ecosystem models can vary from nonspatial,
0-D models, to models that have \mbox{3-D} spatial structure.
\citet{leeds2012modeling} used a 1-D (in the vertical) four-component
model that included nutrients, zooplankton, phytoplankton and detritus
(e.g., a ``1-D NZPD'' model) with iron limitation (i.e., ``1-D
NPZDFe'') to model a \mbox{3-D} process by creating a forest of 1-D models
(more specifically, a forest of emulators of the 1-D model).
These 1-D models resolve vertical processes, but do not account for
horizontal diffusion and advection.
Variability resulting from horizontal advection and diffusion is
accommodated by putting a spatial GP model on the parameters (inputs)
to the 1-D\break  NPZDFe model.

In certain circumstances, the
forward model emulator is developed simply to learn about the behavior
of the
forward model itself. However, in other situations, the
forward model emulator may be useful in a data assimilation context. In
those circumstances, the emulator may be used in a two-stage approach,
where the emulator is fit ``off-line,'' and then is used inside a
Bayesian hierarchical model in the place of the
forward model itself. \citet{calder2011modeling} and \citet
{leeds2012emulator} developed BHMs using
forward model output as data.

\begin{figure*}[t!]

\includegraphics{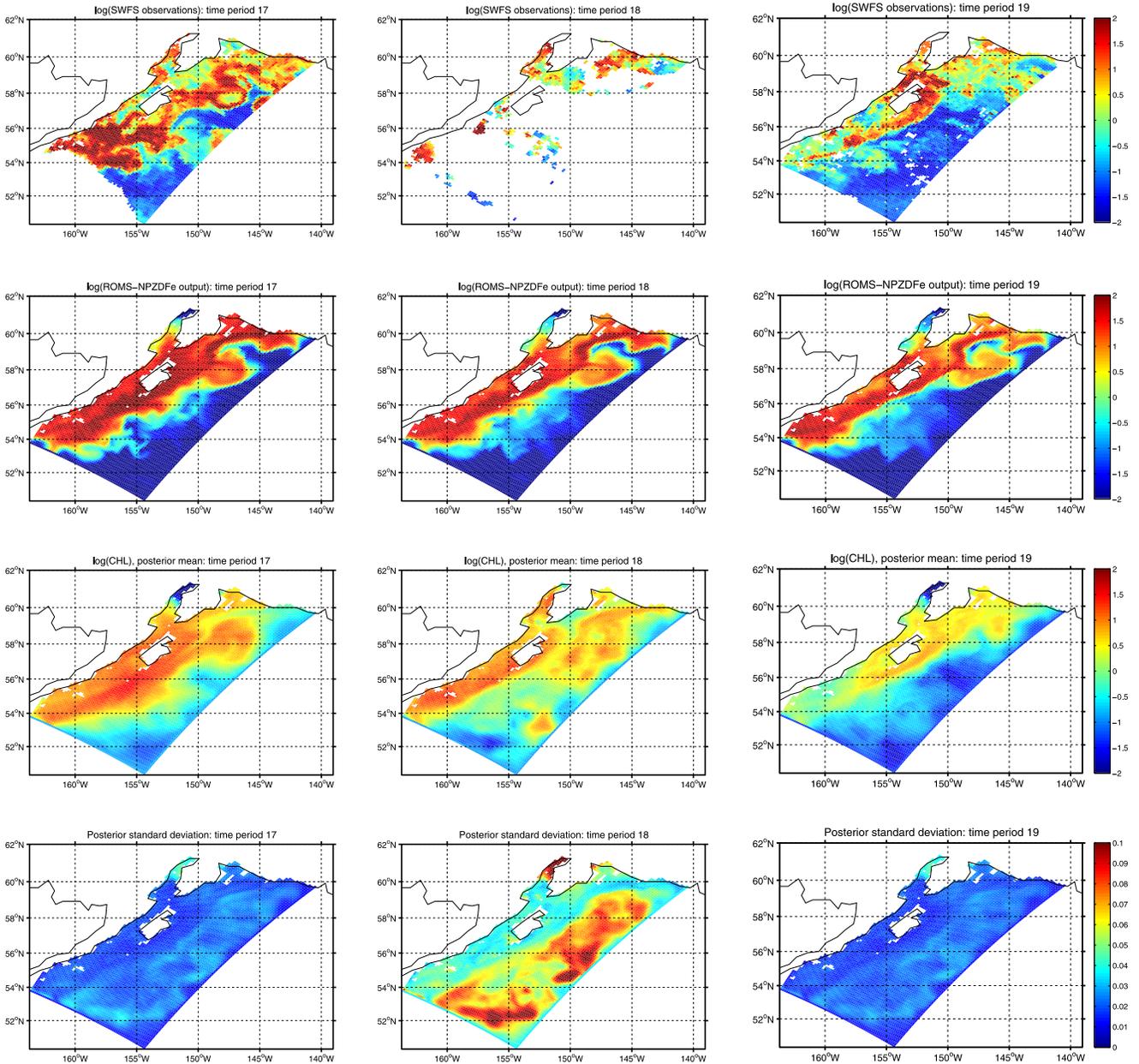}

\caption{A plot of SeaWiFS ocean color observations (top row),
ROMS-NPZDFe output for phytoplankton
(second row), the posterior mean for phytoplankton (third row) and the
posterior standard deviation (bottom row), for three consecutive
eight-day time periods, corresponding to May 16, May 24 and June 1,
2002, respectively. Note that observations were log-transformed and
that the MCMC was run using these log-transformed observations.}
\label{fig.bio}\vspace*{12pt}
\end{figure*}

\citet{leeds2012emulator} performed data assimilation using SeaWiFS
ocean color observations, as well as sea surface height and SST output
from the Regional Ocean Model System (ROMS) forward model coupled with
the NPZDFe model (\cite{FM12}).
They consider a reduced-dimen\-sion quadratic nonlinear process model
similar to (\ref{eq:Yphiphi}) and (\ref{eq:GQNvec}), but where ${\mbf
Y}_{t} = ({\mbf Y}'_{1,t},{\mbf Y}'_{2,t},{\mbf Y}'_{3,t})'$,
representing the three state variables of interest. Basis functions
${\mbv\Phi}^{(1)}$ and ${\mbv\Phi}^{(2)}$ were based on
output from the ROMS-NPZDFe model for a different time period using a
singular value decomposition and the priors for ${\mbf M}_{L}$ and
${\mbf M}_{Q}$ were developed using the remaining right singular
vectors. \citet{leeds2012emulator} show that this approach was
sufficient to accommodate nonlinear dynamics even in the absence of
observations over a substantial portion of the domain as shown in
Figure~\ref{fig.bio}.

\subsection{Other Important Problems}\label{sec3.4}

Space limitations prevent us from giving complete overviews of all
hierarchical statistical models
in\break oceanography. However, we mention here some notable and important
papers that have been published in recent years in areas outside of
those mentioned above. By necessity, this list is not exhaustive,
neither in terms of topics nor citations within topics, but it does
give some sense of the breadth and depth of work being done in the area.

As mentioned above, the amount of {in situ} data for ocean state
variables is fairly limited and there is a need to construct complete
spatial fields representing climatological features of the ocean state,
as well as the uncertainty in those states using spatial and
spatiotemporal models. There are many notable examples from a BHM
perspective (e.g., \cite{higdon1998process};
\cite{lavine1999markov};
Lemos\break and Sans{\'o},
\citeyear{lemos2009spatio,lemos2012conditionally};
\citeauthor{lemos2009hierarchical}\break  (\citeyear{lemos2009hierarchical})).

It has long been known that there are strong teleconnections between
the ocean and the atmosphere and this has led to useful statistical
models for prediction (e.g., \cite{davis1976predictability};
\cite{barnett1981statistical};
\cite{barnett1987origins}).
There is an increasingly large number of papers that have used the BHM
perspective to model
teleconnections as well as their impacts. For example, \citet
{wikle2003climatological} used such a model when evaluating the impacts
of SST on tornado report counts in the eastern two-thirds of the US;
\citet{elsner2006forecasting} (and references therein) have shown great
success in modeling hurricane activity and forecasts (see
also \cite{flay2007effect}); \citet{lima2009hierarchical} use a BHM to model
precipitation considering ocean conditions. In addition, there are
recent studies showing linkages between the physical ocean and
ecological impacts (e.g., \cite{cloern2010biological}; \cite{ruiz2009bayesian}).

Bayesian hierarchical models have long been used to model the higher
trophic levels of the marine ecosystem. For example, stock-recruitment
models have been an important tool in marine fisheries management (e.g.,
\cite{thompson1992bayesian};
\cite{hilborn1994bayesian};
\citeauthor{mcallister1998bayesian}\break
(\citeyear{mcallister1998bayesian});
\cite{dorn2002advice};
\citeauthor{michielsens2004bayesian}\break
(\citeyear{michielsens2004bayesian});
\cite{hirst2005estimating}). In addition,
BHMs have been used extensively in recent years to model distributions
and movement of marine mammals (e.g., \cite{jonsen2007identifying};
Ver~Hoef and\break Jansen (\citeyear{ver2007space});
\cite{johnson2008continuous};
\cite{cressie2009accounting};
\cite{hanks2011velocity};
\cite{moore2011bayesian};
\cite{conn2012accounting};
\cite{hiruki2012bayesian};
\cite{mcclintock2012general}).

Clearly, a critical problem of vast societal interest concerns climate
change. Given the
role of the ocean in the climate system and its direct connections to
weather events and ecological impacts, oceanic climate change and its
uncertainty characterization are extremely important. This is a vast
research area with
many papers that have taken a hierarchical Bayesian approach. This
topic is beyond the scope of this review, but a few notable examples
include \citet{tebaldi2005quantifying}, \citet{furrer2007multivariate},
\citet{tebaldi2008joint}, \citet{aldrin2012bayesian} and \citet
{satterthwaite2012bayesian}.

\section{Current and Future Challenges}\label{sec4}

As demonstrated throughout this review, the ocean system is quite
complex with numerous interacting subcomponents and external processes,
and with substantial uncertainty in terms of knowledge and data.
Statistical methods have been critically important to improve our
understanding of this system and to characterize uncertainty. In recent
years, the hierarchical statistical modeling paradigm has proven to be
an exceptionally useful tool to manage the various sources of
uncertainty. We have only just scratched the surface in terms of our
presentation of the work that has been done in this area, but it is
clear that the use of hierarchical modeling in oceanography has
blossomed. That being said, in addition to continued development in the
areas mentioned above, there are still many important problems at the
interface of Statistics and Oceanography to be considered and
methodologies to be explored.

One of the biggest challenges for statisticians working in Oceanography
is to develop statistical models that can effectively parameterize the
complex nonlinear interactions that are associated with the ocean
system. In particular, models must be developed to account for
potentially high-dimensional state-processes, and yet effectively
manage uncertainty with varying data quality and non-Gaussian errors.
In general, to be effective in this context, dimension reduction should
not be independent of the physical and biological environment of the
problem under consideration.

One component of nonlinearity that is important concerns the coupling
of subsystems, whether that be the atmosphere/ocean interface, the
ocean/ice interface, the physical/biological interface or interactions
between trophic levels in the ocean. Many of these couplings require
parameterizations for fluxes across boundaries, and there is often only
limited observational data available to help inform the process. This
presents an extreme challenge and one in which we must use the large
amounts of remotely sensed observations along with the sparse {in
situ} data to build these nonlinear relationships.

Not unrelated
is the notion of accounting for and describing model error.
The precise characterization of uncertainty that is fundamental in BHM
can be used
to help identify and characterize model error in deterministic models
of the ocean system, for example, ocean forecast models. As with all
model abstractions, deterministic approaches
accept trade-offs in resolution and approximations of the ocean
variability to gain affordability
in simulations and forecasts. Because the ocean system and the models
are inherently nonlinear, the errors
introduced by acceptable approximations can grow and lead to model
error properties that
are difficult to diagnose. Parameters of posterior distributions from
independent BHM analyses
for ocean model response or control variables can be used as a standard
against which deterministic
model error can be quantified. For example, if the incremental
adjustment of the surface
momentum flux control vector in a variational data assimilation
procedure pushes the momentum
flux at a point outside the reasonable bounds of a posterior
distribution for momentum
flux from a BHM, the variational procedure is probably compensating for
forecast model error as opposed
to uncertainty in the control vector.

To gain understanding of the impact of physical processes on biota as
well as the interaction of biological organisms, there is increasing
reliance on individual (or agent-based) models. These models have
become more prevalent in the ecological realm and are starting to be
considered from the BHM perspective (e.g., \cite{hooten2010statistical}). The use of these models in a
statistical context across the spectrum of scales and processes
involved in oceanography is a growing area of interest. For example,
\citet{megrey2007bioenergetics} link physical ocean models to
individual-based bioenergetic models for fish.

The issues described above will almost certainly require new
computational strategies, particularly in the Bayesian paradigm.
As the models get more and more complex and larger data sets become
available, standard MCMC methods begin to fail to provide meaningful
results. Among the many challenges associated with Bayesian computing,
two stand out: (1) there are very high-dimensional distributions to be
explored, and (2) the models are complex, which leads to inexact and
sometimes impossible likelihood evaluations. Novel MCMC methodology
will address these issues, while maintaining feasibility. For example,
particle MCMC (PMCMC) methods are designed to address the first issue.
For high-dimensional posterior distributions (thousands of\break state
variables and parameters), it is nearly impossible to design good
Metropolis--Hastings proposal distributions and in this case, even
adaptive MCMC may fail. \citet{andr2010} propose to use sequential Monte
Carlo (SMC) methods combined with importance sampling to design near
optimal proposal distributions. The resulting algorithm, which has
similar features to a particle filter, will update the entire
collection of state variables at once and can be extremely useful for
space--time models. \citet{parslow2013bayesian} use PMCMC for state and
parameter estimation as well as state forecasting in the context of a
marine biogeochemical model. In addition, Hamiltonian MCMC
methodologies may provide an attractive alternative for situations for
which samples from complicated high-dimensional processes and parameter
distributions are required (e.g., \cite{Beskos2013}).

Approximate Bayes Computing (ABC) methods are a new and emerging class
of likelihood-free\break MCMC algorithms. They address the second issue above---that is,
cases when evaluation of the likelihood function involves
integrals over very large spaces that are impossible to calculate (e.g., \cite{beau2003}). ABC relies on the ability to simulate from the
selected model without much computing effort. Consequently, the user is
forced to select a relevant (multivariate) statistic and ``posterior
samples'' are defined as parameter values that result in a statistic
similar to the one observed. This algorithm, in fact, explores the
distribution of the variables of interest \textit{conditional on the
selected statistic}, not the data. Although this may raise some concern
regarding the interpretation of the results, \citet{fern2012} describe a
semi-automatic way of selecting good summary statistics in a general setting.

In general, it seems likely that for inference and prediction of
complicated oceanographic processes in the foreseeable future, one will
have to continue to make compromises between model complexity and
computational feasibility. Indeed, one must be willing to accept some
reasonable lack of ``optimality'' in the solution in order to make
headway on many of these problems. This is true regardless of whether
one takes a Bayesian or frequentist perspective. The hierarchical
paradigm helps to some extent as it allows one to consider trade-offs
between approximate computational strategies, incorporation of
scientific information and model specification for the different model
components (data, process and parameters) separately.

In conclusion, there is a long history of activity at the interface of
Statistics and Oceanography. In recent years, the hierarchical
statistical paradigm has proven to be very helpful for managing the
uncertainty associated with data, process and parameters in modeling
the ocean and its related systems. There are increasingly more
oceanographers with advanced statistical training and more
statisticians with oceanographic backgrounds and this is sure to lead
to even more innovative and exciting methodological developments in
years to come.

\section*{Acknowledgments}
The authors wish to thank the Editor, AE and reviewer who provided
substantial suggestions to improve the manuscript. In addition, Wikle
acknowledges the support of National Science Foundation (NSF) Grants
DMS-10-49093, OCE-0814934 and Office of Naval Research (ONR) Grant
ONR-N00014-10-0518. Herbei acknowledges support from NSF\break Grant
DMS-12-09142 and ONR Grant N00014-10-1-0488. Milliff acknowledges early
support for BHM research from the Physical Oceanography Program, ONR;
from the NASA, International Ocean Vector Winds Science Team, and
recent support from US Globec under Grant number NSF OCE-0815030. It is
a pleasure to acknowledge our BHM mentor L. Mark Berliner and the
inspirations he provides dating back to his days as Director,
Geophysical Statistics Project at NCAR through the present. We are also
pleased to acknowledge our colleagues and collaborators who participate
in vigorous cross-disciplinary discussions at summer Confabs in
Boulder, CO. In recent years, these include, Professors Berliner, Mevin
B. Hooten, Andrew M. Moore, Nadia Pinardi and Thomas M. Powell, and
Doctors Jeremiah Brown, Manny Fiadeiro and Jerome Fiech\-ter.

%



\end{document}